\documentclass[preprintnumbers,secnumarabic,amsmath,amssymb,nofootinbib,floatfix]{revtex4}
\usepackage{varioref,exscale,latexsym,amsmath,amssymb}
\usepackage{graphicx}

\usepackage{graphicx}
\usepackage{slashed}
\usepackage{dcolumn}
\usepackage{bm}

\def\beq{\begin{equation}}

\def\eeq{\end{equation}}

\def\beqa{\begin{eqnarray}}

\def\eeqa{\end{eqnarray}}

\begin{document}

\title{{\bf On the running of the gravitational constant }}

\medskip\
\author{Mohamed M. Anber${}^{1}$}
\email[Email: ]{manber@physics.utoronto.ca}
\author{ John F. Donoghue${}^{2,3}$}
\email[Email: ]{donoghue@physics.umass.edu}
\affiliation{~\\
${}^{1}$Department of Physics,
University of Toronto\\
Toronto, ON, M5S1A7, Canada\\ \\
${}^{2}$Department of Physics,
University of Massachusetts\\
Amherst, MA  01003, USA\\ \\
${}^3$Niels Bohr International Academy and Discovery Center\\
Niels Bohr Institute\\
Blegdamsvej 17\\
DK-2100 Copenhagen\\
Denmark }

\begin{abstract}
We show that there is no useful and universal definition of a running gravitational constant, G(E), in the perturbative regime below the Planck scale. By consideration of the loop corrections to several physical processes, we show that the quantum corrections vary greatly, in both magnitude and sign, and do not exhibit the required properties of a running coupling constant. We comment on the potential challenges of these results for the Asymptotic Safety program.
\end{abstract}
\maketitle


\section{Introduction}

The concepts of running coupling constants and the renormalization group are
of great utility in renormalizable field theories. It is tempting to attempt a
definition of a running gravitational constant $G(q^2)$ in the context of quantum
general relativity. As a simple example, one could take the quantum correction to the
gravitational potential between heavy masses \cite{potential}\footnote{We do not display the classical correction
to the potential, as it is not relevant for our discussion. For an early attempt to calculate the quantum correction to the gravitational potential see \cite{radkowski}.}
\begin{equation}
V(r) = -G\frac{Mm}{r}\left[1+ \frac{41}{10\pi}\frac{G}{r^2}\right]\,,
\end{equation}
and turn it into a running coupling
\begin{equation}
G(r) = G\left[1+ \frac{41}{10\pi}\frac{G}{r^2}\right]
\end{equation}
which would imply that the gravitational strength gets stronger at short distance. We will
return to the weaknesses of this particular definition in Sec. 8 below, but it is only
one of many attempted definitions that have been suggested in the literature \cite{runningG}.

Moreover there is an extensive approach to quantum gravity that relies on the running of the gravitational
coupling. Within the hypothesis of Asymptotic Safety \cite{Weinberg:1980gg, asymsafety}, a suitably normalized version of the gravitational coupling is proposed to run to an ultraviolet fixed point. In Euclidean space, one treats the dimensionless
combination $g= Gk_E^2$, where $k_E$ is a measure of the Euclidean energy, and the fixed point is described by
\begin{equation}
g= Gk_E^2 \to g_*   ~~~~{\rm as} ~~~k_E \to \infty\,.
\end{equation}
This would imply that the running gravitational strength itself vanishes at large energy, $G(k_E)\to 0$.
This behavior is often summarized by a function
\begin{equation}
G(k_E) = \frac{G_N}{1+\alpha G_N k_E^2}\,,
\label{runningform}
\end{equation}
where $G_N$ is the gravitational coupling at zero energy.

Gravitational corrections are calculable at low energy using effective field
theory \cite{Donoghue:1994dn}. However the effective field theory presents a very different
picture of quantum effects - one in which the gravitational constant does {\em not}
run. Because of the dimensional coupling, loops do not lead to a renormalization
of the leading gravitational action $R$, i.e. the one that contains the gravitational constant $G$,
but rather renormalize higher order terms in the action that come with higher derivatives, i.e.
terms such as $R^2$ or $R_{\mu\nu}R^{\mu\nu}$. Indeed there is a power counting theorem that
says that each gravitational loop always brings in two more powers of the energy.
The content of the renormalization group within effective field theory has been
explored by Weinberg and others \cite{weinberg}. While the technique is useful at
predicting some leading and subleading logarithmic kinematic dependence, it does not
involve the running of the leading order coupling, and in gravity does not
predict the running of $G$. Efforts to define a running coupling in the perturbative regime
must attempt something outside of the usual application of the renormalization group.

The purpose of this paper is to explore explicit physical calculations in Lorentzian spacetime
to see if there is a definition of running gravitational coupling which is physically useful and universal
in the regime where we can maintain control of our calculations. Our answer is negative - no
such definition is both useful and universal.

\section{Gravitational corrections to coupling constants}

Ultraviolet divergences in field theories correspond to counterterms in a local Lagrangian. In
a theory with a dimensional coupling constant with an inverse mass dimension, such as general relativity, these local terms are ones which have a higher mass dimension than the original starting Lagrangian. For example, for gravitational corrections to Quantum Electrodynamics (QED) the divergences generated at one loop have the forms
\begin{equation}
{\cal L}_{ct} = Gc_1 F^{\mu\nu} \partial^2F_{\mu\nu} + Gc_2\bar{\psi}\gamma_\mu \psi \partial_\nu F^{\mu\nu} + Gc_3 \bar{\psi}\gamma_\mu D_\nu \psi F^{\mu\nu}\,,
\end{equation}
when regularized dimensionally. (We will discuss regularization with a dimensionful cutoff in the next section.)
Equivalently field redefinitions and/or equations of motion can be used to rewrite some of these operators (and the accompanying divergences) as contact interactions such as
\begin{equation}
{\cal L'}_{ct} =  Gc_4\bar{\psi}\gamma_\mu \psi \bar{\psi}\gamma^\mu \psi\,.
\end{equation}
The divergences can be absorbed into renormalized parameters of these higher dimensional operators, as in standard effective field theory practice. The scale dependence that arises in dimensional regularization will then be associated with the coefficients of the higher order operators, i.e. the set $c_i$, rather than the original coupling constant, which in the QED case would be the electric charge. Finite remainders will also carry this higher order momentum dependence, i.e. they will be of order $Gq^2$ or $Gq^2\ln q^2$ higher than the original coupling.

Because loops generate these higher order effects in the momentum expansion, attempts to have the loops contribute to the running of the original charge $e$ require some repackaging of the higher order effects into a revised definition of the original charge. Logically this is conceivable by renormalizing the coupling at a higher energy scale. The standard effective field theory treatment is equivalent to renormalzing the operators near zero energy. However, by choosing a renomalization condition that defines the charge at a higher energy scale $E$, one would in general include some of the higher momentum dependence into the definition of the charge.
For example, for a given amplitude $Amp_i$ we would define
\begin{eqnarray}
Amp_i &=& a_i g^2 +b_ig^2\kappa^2 q^2 \nonumber \\
&=& a_i g^2 (1+ \frac{b_i}{a_i}\kappa^2 E^2) + b_i g^2\kappa^2(q^2-E^2)\nonumber \\
&=& a_i g^2(E) + b_ig^2(E) \kappa^2(q^2-E^2)\,,
\end{eqnarray}
when renormalizing at the scale $q^2 =+E^2$. Here $G=\kappa^2/(32\pi)$ and $a_i~b_i$ are process dependent constants.
This  can certainly be done by explicit construction - indeed it can be done in multiple ways through different choices of the renormalization condition. By construction, this provides a gravitational contribution to the running of the coupling with energy.

However there are obstructions that will in general keep any such construction from being useful and universal in effective field theories such as general relativity. One is a kinematic ambiguity. The higher order corrections are proportional to $Gq^2$, and $q^2$ refers to a four-vector which can take both positive and negative values. A renormalization condition defined at one sign of $q^2$ produces a charge definition that fails to be applicable in the crossed reaction with the opposite sign of $q^2$. We will refer to this as the {\em crossing problem}. (In the above schematic example, this is visible in the ambiguity between renormalizing at $q^2=+E^2$ or $q^2=-E^2$.) The other obstruction is process dependence or {\em non-universality}. Because there are in general multiple higher order operators that are possible, and because these enter into different processes in different ways, the divergences and finite parts of different reactions will not be the same. A definition of the charge that is appropriate for one reaction will not work for another. To be useful, a running coupling must capture at least a significant or common portion of the quantum corrections. (In the preceding schematic example, problems would arise if the ratio $b_i/a_i$ is highly process dependent.) We have demonstrated these obstacles in Yukawa theories \cite{Anber:2010uj}, and the same problems arise in the attempts to define gravitational corrections to running gauge charges.

In renormalizable theories these obstacles are not present. For logarithmic running couplings, the kinematic ambiguity is absent because the real part of $\ln q^2$ is the same for both spacelike and timelike $q^2$. The process dependence is absent because the running is connected with the actual renormalization of the charge - the $\ln (q^2/\mu^2)$ factor that arises is tied to the $1/\epsilon$ divergence absorbed into the renormalized charge. Because charge renormalization is universal, the corresponding logarithmic corrections are also universal.

In \cite{Anber:2010uj}, we did find one case where a gravitational correction to a running coupling could be constructed perturbatively without obvious flaws. This was  $\lambda \phi^4$ theory. In this case, the unique higher order operator vanishes by the equation of motion - the gravitational corrections are one-loop-finite. In addition, the complete permutation symmetry of the $\phi^4$ interaction means that all reactions are crossing symmetric, and involve a unique crossing symmetric combination of kinematic invariants. There is no crossing problem. Because there is only one type of vertex in this theory, there is also no non-universality problem. These allowed a reasonable definition of the gravitational correction to the running of $\lambda$. However, we will see that these nice properties are not shared by the gravitational self interactions.

\section{Dimensional cutoffs}

The comments of the preceding section are reasonably obvious when dimensional regularization is used. Indeed they are consistent with the portion of the literature concerning gravitational corrections to gauge couplings that employed dimensional regularization \cite{doesnot}. However, more recently there has been another subset of this literature which used dimensionful cutoffs in the analysis and which reached the opposite conclusion, i.e. that there was a universal gravitational correction to the running of the gauge charges \cite{does, does2}. This dichotomy appears to violate a key principle that true physics is independent of the renormalization scheme. In fact, there is no disagreement between the schemes and the apparent difference arises from an incorrect interpretation of the dimensionful cutoff schemes. It is important for us to demonstrate the flaw of these cutoff calculations, because many attempts to define a running gravitational constant, $G(\Lambda)$ employ a similarly incorrect reasoning.

If we rescale the vector field, $A^\mu \to A^\mu/e_0$, then the electric charge appears only in the photon part of the Lagrangian
\begin{equation}
{\cal L} = -\frac{1}{4e_0^2}F_{\mu\nu}F^{\mu\nu} + \bar{\psi} i \slashed{D}\psi\,.
\end{equation}
After including gravition loops regularized with a dimensional cutoff $\Lambda$, there is a quadratic cutoff dependence in the leading term, as
well as logarithmic cutoff dependence with a high order operator
\begin{equation}
{\cal L} = -\frac{1+a\kappa^2 \Lambda^2}{4e_0^2}F_{\mu\nu}F^{\mu\nu} + b \ln \Lambda^2 ~F_{\mu\nu}\partial^2 F^{\mu\nu} \,.
\label{cutoffL}
\end{equation}
This demonstrates that, in contrast with dimensional regularization, the lowest order charge $e_0$ does get renormalized (quadratically) by graviton loops when using a dimensionful cutoff as a regularizer.

The incorrect interpretation of this is to identify the cutoff dependence with the running of the coupling. Specifically, by writing
\begin{equation}
e^2(\Lambda) =\frac{ e_0^2}{1+ a \kappa^2 \Lambda^2}\,,
\end{equation}
the authors of \cite{does2} identify a beta function
\begin{equation}
\beta (e) = \Lambda \frac{\partial e}{\partial \Lambda} =-  a e\kappa^2 \Lambda^2
\end{equation}
for the running of the coupling.

However this interpretation is incorrect, because the quadratic $\Lambda$ dependence disappears from physical observables
in the process of renormalization. For example, the Coulomb potential at low energy, calculated from Eq. \ref{cutoffL} is
\begin{equation}
{ V}(r) = \frac{ e_0^2}{4\pi(1+ a \kappa^2 \Lambda^2)} \frac1{r}\,.
\end{equation}
If we use this to identify the electric charge we obtain
\begin{equation}
\alpha = \frac{e^2}{4\pi} =\frac{ e_0^2}{4\pi(1+ a \kappa^2 \Lambda^2)}  =\frac1{137}\,.
\end{equation}
When expressing predictions in terms of the measured value of $\alpha$, the quadratic $\Lambda$ dependence is removed from
all observables at all energies. It does not indicate the running of the electric charge. This analysis is supported by explicit calculation \cite{Toms:2011zz}.

These same comments apply to dimensionful cutoffs and the gravitational coupling $G$. When using a scheme with a dimensionful cutoff, one
will generate corrections to the gravitational constant $G=G_0(1+aG_0\Lambda^2)$. If this is done in a way that preserves general covariance,
the same correction will be obtained in any process that involves $G$. This quadratic dependence will disappear from all observables once one
identifies the physical renormalized parameter $G$ to be equal to its Newtonian value.

Note that the logarithmic $\ln \Lambda$ dependence can be useful in tracing the running of couplings. This is because at high energies the logarithm must also contain kinematic variables, i.e. $\ln (\Lambda^2/q^2)$. This is analogous to tracing the $\ln q^2 $ behavior from the $1/\epsilon$ dependence in dimensional regularization.

These features explain how dimensional regularization and cutoff regularization can agree in calculations involving gravity. There are no quadratic divergences in dimensional regularization, but we have seen that such divergences in cutoff schemes disappear from observables under renormalization. When considering gravitational corrections, the logarithmic cutoff dependence and the $1/\epsilon$ dependence are both associated with higher order terms of order $\kappa^2 q^2$, and the residual kinematic effects will be in agreement when the calculations are properly done. Dimensional regularization is a good regulator for the gravitational interaction because it has a clear and direct interpretation. For this reason, we use dimensional regularization throughout this paper.

\section{Pure gravity: The graviton propagator}

Let us first consider the purely gravitational sector. At one loop, pure gravity is finite for on-shell amplitudes, because the
higher order counterterms, $R^2$ and $R_{\mu\nu} R^{\mu\nu}$ vanish by the equation of motion $R_{\mu\nu}=0$. Finite one-loop corrections
do exist. These are higher order in the momentum variables, and do not have the same kinematic dependence as the lowest order effects
governed by the gravitational constant $G$. However, by working at a high energy renormalization scale, we will see if we can attempt to package these loop effects as a running coupling $G(E)$.

Even though our primary focus in this paper involves on-shell physical reactions, it is useful to start by consideration of the vacuum polarization diagram and the graviton propagator. The only quantum correction that is demonstrably universal is that involving the vacuum polarization. Every graviton exchange receives a correction from the vacuum polarization. Because each end of the propagator carries a factor of $\kappa$, a modification
of the propagator could be interpreted as modification of the coupling $G(q^2)$. However, we will see that there still remains the crossing problem because $q^2$ can carry either sign.

The graviton propagator at lowest order is given by
\begin{equation}
iD_F^{\alpha\beta\gamma\delta} = \frac {i{\cal
P}^{\alpha\beta\gamma\delta}}{q^2+i\epsilon}\,,
\end{equation}
where
\begin{equation}
{\cal P}^{\alpha\beta\gamma\delta} =
\frac12\left[\eta^{\alpha\gamma}\eta^{\beta\delta} +
\eta^{\beta\gamma}\eta^{\alpha\delta}
-\eta^{\alpha\beta}\eta^{\gamma\delta}\right]\,.
\end{equation}
The inclusion of the vacuum polarization diagram modifies the propagator
\begin{equation}
{i{\cal
P}^{\alpha\beta\lambda\xi}\over
q^2}i\Pi_{\lambda\xi\mu\nu}(q){i{\cal P}^{\mu\nu\gamma\delta}\over
q^2}~.
\end{equation}

The vacuum polarization diagram is divergent and the required counter terms have the form obtained by `t Hooft and
Veltman \cite{'tHooft:1974bx}
\begin{eqnarray}
\Delta {\cal L}=\frac{\sqrt{g}}{16\pi^2\epsilon}\left[\frac{1}{120}R^2+\frac{7}{20}R_{\alpha\beta}R^{\alpha\beta} \right]
\end{eqnarray}
with $\epsilon=(d-4)/2$. Because such terms appear in the most general effective
Lagrangian, one can absorb these divergences into the renormalized values of
their coefficients. Expressing $R^2$ and $R_{\alpha\beta}R^{\alpha\beta}$ in terms of the Fourier-space momenta, we find upon symmetrizing the indices
\begin{eqnarray}
\sqrt{g}R^2=h^{\mu\nu}\left[ q^4\eta_{\alpha\beta}\eta_{\mu\nu}-q^2\left(\eta_{\mu\nu} q_\alpha q_\beta+ \eta_{\alpha\beta} q_\mu q_\nu \right) +q_{\alpha}q_{\beta}q_{\mu}q_{\nu} \right]h^{\alpha\beta}\,,
\end{eqnarray}
and
\begin{eqnarray}
\nonumber
\sqrt{g}R_{\alpha\beta}R^{\alpha\beta}&=&\frac{1}{4}h^{\mu\nu}\left[q^4\eta_{\alpha\beta}\eta_{\mu\nu}-\frac{q^2}{2}\left(q_{\alpha}q_\mu\eta_{\nu\beta}+q_{\alpha}q_{\nu}\eta_{\mu\beta}+q_{\beta}q_{\mu}\eta_{\nu\alpha}+q_{\beta}q_{\nu}\eta_{\mu\alpha}\right) \right.\\
&&\left.\quad\quad\quad -q^2(\eta_{\alpha\beta}q_\mu q_\nu+\eta_{\mu\nu}q_\alpha q_\beta)+q^4I_{\alpha\beta,\mu\nu}+2q_\alpha q_\beta q_\mu q_\nu\right]h^{\alpha\beta}\,,
\end{eqnarray}
where $I_{\alpha\beta,\mu\nu}=(\eta_{\alpha\mu}\eta_{\beta\nu}+\eta_{\alpha\nu}\eta_{\beta\mu})/2$.
 In addition, one can use the presence of the
$1/\epsilon$ terms to read out the dependence on $\ln q^2$
\begin{eqnarray}
\nonumber
\Pi_{\alpha\beta,\mu\nu}(q)&=&-\frac{2G}{\pi}\ln\left(-\frac{q^2}{\mu_1^2} \right) \left[\frac{q^4}{60}\eta_{\alpha\beta}\eta_{\mu\nu}-\frac{q^2}{60}\left(\eta_{\mu\nu} q_\alpha q_\beta+ \eta_{\alpha\beta} q_\mu q_\nu \right) +\frac{1}{60}q_{\alpha}q_{\beta}q_{\mu}q_{\nu} \right]\\
\nonumber
&&-\frac{2G}{\pi}\ln\left(-\frac{q^2}{\mu_2^2}\right)\left[\frac{7}{40}q^4\eta_{\alpha\beta}\eta_{\mu\nu}-\frac{7}{80}q^2\left(q_{\alpha}q_\mu\eta_{\nu\beta}+q_{\alpha}q_{\nu}\eta_{\mu\beta}+q_{\beta}q_{\mu}\eta_{\nu\alpha}+q_{\beta}q_{\nu}\eta_{\mu\alpha}\right) \right.\\
&&\left.\quad\quad\quad\quad\quad\quad\quad\quad -\frac{7}{40}q^2(\eta_{\alpha\beta}q_\mu q_\nu+\eta_{\mu\nu}q_\alpha q_\beta)+\frac{7}{40}q^4I_{\alpha\beta,\mu\nu}+\frac{7}{20}q_\alpha q_\beta q_\mu q_\nu\right]\,,
\end{eqnarray}
where we have assigned $\ln(-q^2/\mu_1^2)$ and $\ln(-q^2/\mu_2^2)$  for $R^2$ and $R_{\alpha\beta}R^{\alpha\beta}$ respectively.

The polarization tensor can be written in the form \cite{Weinberg:1980gg}
\begin{eqnarray}
\Pi^{\alpha\beta,\mu\nu}(q)&=&q^4A(q^2)L^{\alpha\beta}(q)L^{\mu\nu}(q)-q^2B(q^2)\left[L^{\alpha\mu}(q)L^{\beta\nu}(q)+L^{\alpha\nu}(q)L^{\beta\mu}(q) -2L^{\alpha\beta}(q)L^{\mu\nu}(q)\right],
\end{eqnarray}
where $L^{\mu\nu}(q)=\eta^{\mu\nu}-q^{\mu}q^{\nu}/q^2$.
Contracting $\Pi^{\alpha\beta,\gamma\delta}(q)$ with $L^{\alpha\beta}(q)L^{\mu\nu}(q)$, and $L^{\alpha\mu}(q)L^{\beta\nu}(q)$ we obtain two equations in $A(q^2)$, and $B(q^2)$
\begin{eqnarray}
\nonumber
\Pi^{\alpha\beta,\mu\nu}(q)L_{\alpha\beta}(q)L_{\mu\nu}(q)&=&3q^2\left[3q^2A(q^2)+4B(q^2) \right]\,,\\
\Pi^{\alpha\beta,\mu\nu}(q)L_{\alpha\mu}(q)L_{\beta\nu}(q)&=&3q^2\left[q^2A(q^2)-2B(q^2) \right]\,.
\end{eqnarray}
Hence, we find
\begin{eqnarray}
\nonumber
A(q^2)&=&-\frac{1}{30\pi}G\ln\left(\frac{-q^2}{\mu_1^2}\right)-\frac{7}{10\pi}G\ln\left(\frac{-q^2}{\mu_2^2}\right)\,,\\
B(q^2)&=&\frac{7}{40\pi}Gq^2\ln\left(\frac{-q^2}{\mu_2^2}\right)\,.
\end{eqnarray}

On the other hand, the bare propagator takes the general form
\begin{eqnarray}
i{\cal D}^{\alpha\beta,\mu\nu}(q^2)=\frac{i}{2q^2}\left[L^{\alpha\mu}L^{\beta\nu}+L^{\alpha\nu}L^{\beta\mu}-L^{\alpha\beta}L^{\mu\nu} \right]\,,
\end{eqnarray}
while the quantum corrected propagator reads
\begin{eqnarray}
\nonumber
i{\cal D}'^{\alpha\beta,\mu\nu}&=&i{\cal D}^{\alpha\beta,\mu\nu}+i{\cal D}^{\alpha\beta,\gamma\delta}i\Pi_{\gamma\delta,\rho\tau}i{\cal D}^{\rho\tau,\mu\nu}\\
&=&\frac{i}{2q^2}(1+2B(q^2)) \left[L^{\alpha\mu}L^{\beta\nu}+L^{\alpha\nu}L^{\beta\mu}-L^{\alpha\beta}L^{\mu\nu} \right]-i\frac{A(q^2)}{4}L^{\alpha\beta}L^{\mu\nu}\,.
\label{dressed}
\end{eqnarray}
The first term above is a dressed propagator. Therefore, it is appropriate to define the running coupling as
\begin{eqnarray}
\label{one way of normalizing}
 G(q^2)=G(1+2B(q^2))=G\left(1+\frac{7}{20\pi}Gq^2\ln\left(\frac{-q^2}{\mu_2^2}\right)\right)\,.
\end{eqnarray}
On the other hand, the second term in Eq. \ref{dressed} also contributes comparably to gravitational amplitudes. For example, non-relativistic scattering involves the $00,00$ component of ${\cal D}^{\mu\nu,\alpha\beta}$ and we can equally define a running coupling from that component
\begin{eqnarray}
{\cal D}^{00,00}=\frac{1}{2q^2}\left(1+2B-\frac{q^2A}{2}  \right)L^{00}L^{00}=\frac{1}{2q^2} G(q^2)L^{00}L^{00}\,,
\end{eqnarray}
and hence,
\begin{eqnarray}
\label{0000 components}
G(q^2)=G\left[1+\frac{1}{60\pi} Gq^2\ln\left(\frac{-q^2}{\mu_2^2}\right)+\frac{7}{10\pi} Gq^2\ln\left(\frac{-q^2}{\mu_1^2}\right) \right]\,.
\end{eqnarray}
This corresponds exactly to the vacuum polarization contribution to the shift in the Newtonian interaction as calculated in \cite{Donoghue:1994dn}, but it could be more widely applicable. Since  $M_P$ is the only scale in gravity, we expect $\mu_1$ and $\mu_2$ to be of this order, and hence Eqs. \ref{one way of normalizing} and \ref{0000 components} are valid for low energies $E<M_P$. For spacelike $q^2$ this corresponds to an increase in the strength of gravity, while for timelike values it is a decrease. Of course this sign ambiguity is a signal of the crossing problem. It is present no matter which components of the propagator are considered. If converted to Euclidean space by the Wick rotation $q^2 \to (iq_4)^2 -\mathbf{q}^2 =-q_E^2$, the effective strength increases.

At one loop order, it would be hard to convincingly favor one of the two definitions Eqs. \ref{one way of normalizing} or \ref{0000 components}. However, they do at least have a common sign.

However, we know ahead of time that this definition is not going to enter into any physical one-loop processes in pure gravity. Since on-shell
processes are one loop finite, any reference to the higher order operators $R^2$ or $R_{\mu\nu}R^{\mu\nu}$ must drop out of physical observables at one loop. The coefficients $\mu_1$ and $\mu_2$ are in this category and will not appear in on-shell processes. On-shell reactions may have logarithms such as $\ln s/t$, but not $\ln s/\mu_{1,2}^2$.

\section{Pure gravity: Graviton scattering}

The simplest physical process in pure gravity is graviton-graviton scattering. The lowest order scattering amplitude involves a
large number of individual tree diagrams but is given by the simple form
\begin{equation}
{A}^{tree}(++;++) = i\frac{\kappa^2}{4} \frac{s^3}{tu}\,,
\label{gravtree}
\end{equation}
where the signs $+,-$ refer to helicity indices and $s,t,u$ are the usual Mandelstam variables. In power counting, this is a dimensionless
amplitude of order $GE^2$. Our labeling of momentum and helicities corresponds to the final state particle being outgoing, in contrast to
some conventions in the literature which label all particles as ingoing\footnote{The amplitude in Eq. \ref{gravtree} is often referred to as the maximally helicity violating (MHV) amplitude in the all-ingoing convention.}.

The one loop amplitudes have been calculated by Dunbar and Norridge \cite{Dunbar:1994bn}. These are of order $G^2E^4$ and take the form

\begin{eqnarray}\label{eq:2}
{\cal A}^{1-loop}(++;--) & = & -i\,{\kappa^4 \over 30720
\pi^2}
\left( s^2+t^2 + u^2 \right) \,,  \nonumber\\
{\cal A}^{1-loop}(++;+-) & = & -{1 \over 3}
{\cal A}^{1-loop}(++;--)\nonumber \\
{\cal A}^{1-loop}(++;++) & = &\frac{\kappa^2}
{4(4\pi)^{2-\epsilon}}\,
 \frac{\Gamma^2(1-\epsilon)\Gamma(1+\epsilon)}
 {\Gamma(1-2\epsilon)}\,
 {\cal A}^{tree}(++;++)\,\times(s\,t\,u)\\
&&\hspace{-0em}\times\left[\rule{0pt}{4.5ex}\right.
\frac{2}{\epsilon}\left(
\frac{\ln(-u)}{st}\,+\,\frac{\ln(-t)}{su}\,+\,\frac{\ln(-s)}{tu}
\right)+\,\frac{1}{s^2}\,f\left(\frac{-t}{s},\frac{-u}{s}\right)
\nonumber\\&&\hspace{1.4em}
+2\,\left(\frac{\ln(-u)\ln(-s)}{su}\,+\,\frac{\ln(-t)\ln(-s)}{tu}\,+\,
\frac{\ln(-t)\ln(-s)}{ts}\right)
\left.\rule{0pt}{4.5ex}\right]\,,\nonumber
\end{eqnarray}
where
\begin{eqnarray}\label{eq:f}
f\left(\frac{-t}{s},\frac{-u}{s}\right)&=&
\frac{(t+2u)(2t+u)\left(2t^4+2t^3u-t^2u^2+2tu^3+2u^4\right)}
{s^6}
\left(\ln^2\frac{t}{u}+\pi^2\right)\nonumber\\&&
+\frac{(t-u)\left(341t^4+1609t^3u+2566t^2u^2+1609tu^3+
341u^4\right)}
{30s^5}\ln\frac{t}{u}\nonumber\\&&
+\frac{1922t^4+9143t^3u+14622t^2u^2+9143tu^3+1922u^4}
{180s^4}\,.
\end{eqnarray}
Other amplitudes can be obtained from these by crossing.

The dimensional regularization parameter $\epsilon =(4-d)/2$ appears in the amplitude ${\cal A}^{1-loop}(++;++)$. This
is an infrared divergence, and is canceled as usual from the radiation of soft gravitons.
An explicit calculation of the sum of direct and radiative cross sections \cite{Donoghue:1999qh}
yields the result
\begin{eqnarray}\label{sum-crs}
&&\hspace{-3em}\left(\frac{d\sigma}{d\Omega}\right)_{tree}
 + \left(\frac{d\sigma}{d\Omega}\right)_{rad.}
  +\left(\frac{d\sigma}{d\Omega}\right)_{nonrad.}=\\
&=& \frac{\kappa^4 s^5}{2048\pi^2 t^2 u^2}\,
\left\{ \rule{0pt}{2.6em}\right.
1 + {\kappa^2 s \over 16 \pi^2}
\left[\rule{0pt}{2em}\right.
 \ln\frac{-t}{s}\ln\frac{-u}{s}+
\frac{tu}{2s^2}\,f\left(\frac{-t}{s},\frac{-u}{s}\right)
\nonumber\\
&&-\left(\frac{t}{s}\,\ln{\frac{-t}{s}}+\frac{u}{s}\,
\ln{\frac{-u}{s}}\right)
\left(
3\ln(2\pi^2)+\gamma+\ln\frac{s}{\mu^2}+
  \frac{\sum_{ij}\eta_i\eta_j{\cal
F}^{(1)}(\gamma_{ij})}{\sum_{ij}\eta_i\eta_j{\cal
F}^{(0)}(\gamma_{ij})}
\right)
\left.\rule{0pt}{2em}\right]
 \left.\rule{0pt}{2.6em}\right\}.\nonumber
\end{eqnarray}
Here $\mu$ is an infrared scale related to the experimental energy resolution and ${\cal F}^{(i)}$ are functions defined in \cite{Donoghue:1999qh} related to the angles of emission of soft graviton radiation.

The last line of the cross section formula is related to infrared physics and does not appear appropriate for inclusion in the definition of a running coupling. Instead we focus on the correction displayed in the preceding line. We would like a renormalization point in the physical region\footnote{The scattering amplitudes quoted are only valid on-shell. Off-shell evaluation would involve divergences.} with a single energy scale $E$. We choose the central physical point $s=2E^2,~ t=u=-E^2$. This leads to the identification
\begin{equation}
G^2(E) = G^2\left[1 +\frac{\kappa^2 E^2\left(\ln^2 2+\frac{1}{8}\left(\frac{2297}{180}+\frac{63\pi^2}{64}   \right)  \right)}{8\pi^2}\right]~~~.
\label{running1}
\end{equation}
We see that this definition leads to a growing running coupling $G(E)$, as opposed to the expectation from asymptotic safety of a
decrease in strength at high energy.  It works acceptably for this process because it absorbs the main effects of the quantum corrections
in the neighborhood of the central point.

We could alternatively consider the crossed reaction ${\cal A}(+,-;+,-)$ which is obtained from ${\cal A}(+,+;+,+)$ by the exchange $s\leftrightarrow t$. This makes the quantum corrections somewhat different, with the corresponding kinematic factor being
\begin{equation}
1 + \frac{\kappa^2 t}{16 \pi^2}
\left[ \ln\frac{-s}{t}\ln\frac{-u}{t}+ \frac{su}{2t^2} f \left(\frac{-s}{t},\frac{-u}{t}\right) \right]
= 1 +\frac{\kappa^2E^2\left(\frac{29}{10}\ln 2-\frac{67}{45}  \right)}{16\pi^2}
\end{equation}
instead of the factor in Eq. \ref{running1}. The quantum corrections in this channel differ from  those of the original reaction and they are
not accurately summarized by the same running coupling. This is a manifestation of the crossing problem.

\section{Gravitational scattering of a massless scalar particle}

In renormalizable gauge theories, the running coupling applies universally to all processes. As mentioned above, this is because the running is tied  to the renormalization of the gauge charge. General covariance requires that a valid definition of a running $G$ also be universally applicable. The gravitational coupling not only parameterizes the self interactions of gravitons, but it also describes the gravitational coupling of matter. In this section, we look at the effects of loops on the gravitational interactions of a scalar particle.

We consider a scalar particle that has only gravitational interactions. The scattering $\phi + \phi \to \phi +\phi$ via graviton exchange at tree level has $s,~t, ~{\rm and}~ u$ channel poles, with amplitude
\begin{eqnarray}
{\cal M}_{tree}=i\frac{\kappa^2}{4}\left[\frac{st}{u}+\frac{su}{t}+\frac{tu}{s} \right]\,.
\end{eqnarray}
We note that this amplitude, and the loop amplitudes to follow, has a permutation (crossing) symmetry such that all channels are governed by the same amplitude. This will eliminate the crossing problem that arises in most other reactions.  However we can test for universality by testing whether the interactions lead to a similar running coupling as suggested in the purely gravitational sector.

In this theory there is a higher order operator which is required at one loop. Divergences proportional to
\begin{equation}
{\cal L}_2 = \frac{203}{320\epsilon} (D_\mu \phi D^\mu \phi)^2
\end{equation}
arise at one loop. In matrix elements, this operator generates a contribution proportional to $s^2 + t^2 +u^2$.
The one loop amplitudes, up to rational terms in the kinematic variables,
have been given in \cite{Dunbar:1995ed}.
However the rational terms are constrained by the permutation symmetry to also be proportional to $s^2 + t^2 +u^2$ and
we will absorb them into the higher order Lagrangian ${\cal L}_2$.

The total scattering amplitude of this process, apart from a polynomial in $s$, $t$, and $u$, is given by
\footnote{We correct for a few typos found in the result Eq 4.12 of \cite{Dunbar:1995ed}. We verified Eq. \ref{AAAAScattering} by directly computing the whole set of the one-loop Feynman diagrams. }
\begin{eqnarray}
\nonumber
{\cal M}_{1-loop}&=&i\frac{\kappa^4}{\left(4\pi\right)^2}\left[\frac{(s^4+t^4)}{16}I_4(s,t)+\frac{(s^4+u^4)}{16}I_4(s,u)+\frac{(u^4+t^4)}{16}I_4(t,u)-\frac{s(s^2+2t^2+2u^2)}{8}I_3(s)  \right.\\
\nonumber
&&\left.\quad\quad\quad\quad\quad-\frac{t(t^2+2s^2+2u^2)}{8}I_3(t)-\frac{u(u^2+2t^2+2s^2)}{8}I_3(u)+\frac{(163u^2+163t^2+43tu)}{960}I_2(s)  \right.\\
\label{AAAAScattering}
&&\left.\quad\quad\quad\quad\quad +\frac{(163u^2+163s^2+43us)}{960}I_2(t)+\frac{(163s^2+163t^2+43ts)}{960}I_2(u) \right]\,,
\end{eqnarray}
where the $I_4(s,t)$, $I_3(s)$, and $I_2(s)$ are respectively the scalar box, triangle and bubble integrals:
\begin{eqnarray}
\nonumber
I_4(s,t)&=&\frac{1}{st}\left\{\frac{2}{\epsilon^2}\left[(-s)^{-\epsilon}+(-t)^{-\epsilon} \right]-\ln^2\left(\frac{-s}{-t}\right)-\pi^2  \right\}\\
\nonumber
&&=\frac{1}{st}\left\{\frac{4}{\epsilon^2}-\frac{2\ln(-s)+2\ln(-t)}{\epsilon}+2\ln(-s)\ln(-t)+\mbox{finite} \right\}\,,\\
\nonumber
I_3(s)&=&\frac{1}{\epsilon^2}\left(-s\right)^{-1-\epsilon}=-\frac{1}{s}\left(\frac{1}{\epsilon^2}-\frac{\ln(-s)}{\epsilon}+\frac{\ln^2(-s)}{2} \right)\,,\\
I_2(s)&=&\frac{1}{\epsilon(1-2\epsilon)}\left(-s\right)^{-\epsilon}=\left(\frac{1}{\epsilon}-\ln(-s)+\mbox{finite}\right)\,.
\end{eqnarray}

We follow Ref. \cite{Dunbar:1995ed} in removing the infrared divergences by use of
\begin{equation}
\label{removing IR}
{\cal M}_{IR} =\frac{\kappa^2}{2(4\pi)^2}\frac{\left((-s)^{1-\epsilon}+(-t)^{1-\epsilon}+(-u)^{1-\epsilon}\right)}{\epsilon^2}{\cal M}_{tree}\,,
\end{equation}
where the residual hard part is defined via
\begin{equation}
{\cal M}_h = {\cal M}_{1-loop} - {\cal M}_{IR}\,.
\end{equation}
With the renormalization of the higher order operator, the one loop hard amplitude is
\begin{eqnarray}
\nonumber
{\cal M}_{h}&=&i\frac{\kappa^4}{\left(4\pi\right)^2}\left\{\frac{(s^4+t^4)}{8st}\ln(-s)\ln(-t)+\frac{(s^4+u^4)}{8su}\ln(-s)\ln(-u)+\frac{(u^4+t^4)}{8tu}\ln(-t)\ln(-u) \right.\\
\nonumber
&&\left.\quad\quad\quad\quad+\frac{(s^2+2t^2+2u^2)}{16}\ln^2(-s) +\frac{(t^2+2s^2+2u^2)}{16}\ln^2(-t)+\frac{(u^2+2t^2+2s^2)}{16}\ln^2(-u)  \right.\\
\nonumber
&&\left.\quad\quad\quad\quad+\frac{1}{16}\left(\frac{st}{u}+\frac{tu}{s}+\frac{us}{t}  \right)\left(s\ln^2(-s)+t\ln^2(-t)+u\ln^2(-u)  \right)\right.\\
\nonumber
&&\left.\quad\quad\quad\quad+\left[ -\frac{(163u^2+163t^2+43tu)}{960}\ln\left(\frac{-s}{\mu^2}\right)-\frac{(163u^2+163s^2+43us)}{960}\ln\left(\frac{-t}{\mu^2}\right)\right.\right.\\
\label{AAAAhardScattering}
&&\left.\left.\quad\quad\quad\quad-\frac{(163s^2+163t^2+43ts)}{960}\ln\left(\frac{-u}{\mu^2}\right)+d_1^{ren}(\mu)(s^2+t^2+u^2) \right]\right\}\,,
\end{eqnarray}
where $\mu$ is an infrared scale.
In this result, we have grouped the single logs
with the higher order operator, because those logs are the ones that pick up the scale dependence when you shift the scale associated with the higher order operator $ d_1^{ren}(\mu)$.

We again evaluate the matrix element at the central kinematic point $s=2E^2,~t=u=-E^2$. The result is
\begin{equation}
{\cal M}_{total}={\cal M}_{tree}+{\cal M}_h=  i\frac{9\kappa^2 E^2}{8}\left[1-\frac{\kappa^2E^2}{360\left(4\pi\right)^2} \left(609\ln\frac{E^2}{\mu^2}+\left(340\pi^2+\left(123-340\ln 2\right)\ln2  \right)  \right)   \right]\,.
\end{equation}
 If we were to use this to identify a running coupling the result would be
\begin{equation}
G(E) = G\left[1-\frac{\kappa^2E^2}{360\left(4\pi\right)^2} \left(609\ln\frac{E^2}{\mu^2}+\left(340\pi^2+\left(123-340\ln 2\right)\ln2  \right)  \right)   \right]\,.
\label{running2}
\end{equation}
The single log term which appears in Eq. \ref{running2} could reasonably be associated with the higher order operator $d_1$, and perhaps should be removed from this expression. The most serious flaw of this result in comparison to Eq. \ref{running1} is that it has the opposite sign. The leading corrections to graviton scattering and to scalar scattering go in the opposite direction, which is not accountable by a common definition of a running coupling, an obvious lack of universality.

\section{Gravitational scattering of non-identical particles}

Here we consider a different situation for the matter couplings - the scattering of non-identical particles. We will neglect the particle masses, so this corresponds to scattering at $s>>m^2$. This situation demonstrates
both the crossing problem and the non-universality problem.

 The example of the last section has more crossing symmetry than most gravitational reactions. Processes involving non-identical particles, or with fermions, will typically involve dominantly only one of the $s,t,u$ channels. Typical gravitational scattering of very massive particles will involve primarily $t$-channel exchange. Such distinctions highlight the difficulty of any given definition of a running $G$ being applicable to all processes.

By a direct computation of the appropriate set of Feynman diagrams, we find that the tree and one-loop amplitudes of the reaction $A+B\to A+B$ are
\begin{eqnarray}
\nonumber
{\cal M}_{tree}&=&\frac{i\kappa^2 su}{4t}\,,\\
\nonumber
{\cal M}_{1-loop}&=&i\frac{\kappa^4}{\left(4\pi \right)^2}\left[\frac{1}{16}\left(s^4I_4(s,t)+u^4 I_4(u,t)\right) +\frac{1}{8}\left(s^3+u^3+tsu \right)I_3(t)-\frac{1}{8}\left(s^3I_3(s)+u^3I_3(u)  \right)  \right.\\
\label{ABAB scattering}
&&\left.\quad\quad\quad-\frac{1}{240}\left(71us-11t^2\right)I_2(t)+\frac{1}{16}\left(s^2I_2(s)+u^2I_2(u)\right)  \right]\,.
\end{eqnarray}
Then, we use Eq. \ref{removing IR} in removing the IR divergences. The resulting hard amplitude reads
\begin{eqnarray}
\nonumber
{\cal M}_{h}&=&i\frac{\kappa^4}{\left(4\pi \right)^2}\left[\frac{1}{8}\left(\frac{s^3}{t}\ln(-s)\ln(-t)+\frac{u^3}{t} \ln(-u)\ln(-t)\right) -\frac{1}{16t}\left(s^3+u^3+tsu \right)\ln(-t)+\frac{1}{16}\left(s^2\ln^2(-s)+u^2\ln^2(-u)\right)     \right.\\
\nonumber
&&\left.\quad\quad\quad+\frac{us}{16t}\left(s\ln^2(-s)+t\ln^2(-t)+u\ln^2(-u)  \right)+\frac{1}{240}\left(71us-11t^2\right)\ln(-t)-\frac{1}{16}\left(s^2\ln(-s)+u^2\ln(-u)\right) \right]\,,\\
\end{eqnarray}
and the total amplitude at the center kinematic  point $s=2E^2,~t=u=-E^2$ is
\begin{eqnarray}
{\cal M}_{total}=\frac{i\kappa^2E^2}{2}\left[1-\frac{\kappa^2 E^2}{10(4\pi)^2}\left(\left(19+10\ln 2\right)\ln\left(\frac{E^2}{\mu^2}\right)+5\left(\pi^2-(\ln 2-1)\ln 2  \right) \right)  \right]\,.
\end{eqnarray}

On the other hand, the amplitude of the reaction $A+A\to B+B$ is given by Eq. \ref{ABAB scattering} with the exchange $s\leftrightarrow t$, and  has the amplitude
\begin{eqnarray}
{\cal M}_{total}=\frac{i\kappa^2E^2}{8}\left[1+\frac{\kappa^2 E^2}{10(4\pi)^2}\left(9\ln\left(\frac{E^2}{\mu^2}\right)-5\pi^2+\left(19+5\ln 2  \right)\ln2 \right)  \right]\,.
\end{eqnarray}
The crossing problem is obvious here. The loop corrections are in opposite directions in the two reactions, largely because of the change in sign of the kinematic variables under crossing. Any definition of a running $G$ cannot capture this behavior - the coupling must either increase with energy scale or decrease with energy. The processes also illustrate the non-universality problem. Even two reactions that are this closely related have a different magnitude for the one loop correction when evaluated in the physical region.

\section{Gravitational scattering of heavy masses}

Finally we consider the quantum corrections to the scattering of heavy objects - let us call them planets. Are the gravitational corrections here similar to those of massless scalars or gravitons? This tests the universality property of a running coupling. This scattering amplitude is closest to the situations that we are familiar with defining the running couplings in QED or QCD

The gravitational interaction at one-loop is the result of several Feynman diagrams that vary in magnitude and sign. Written in coordinate space, the
total loop correction changes the interaction to that quoted above in Eq. 1. Written in its original momentum space for this corresponds to
\begin{equation}
V(q) = -4\pi\frac{GMm}{\mathbf{q}^2}\left[1+\frac{41}{20\pi}G{\mathbf{q}}^2\ln \left( \frac{\mu^2}{{\mathbf q}^2}\right)\right]\,.
\label{potential}
\end{equation}
Note that in this case, we have written this result in terms of the spatial part of $q^2$, i.e. $q^2=-{\mathbf q}^2$ because the results were derived in the non-relativistic approximation. We should not consider crossing this amplitude to timelike $q^2$ because the planet masses could be well above the Planck scale. This result corresponds to an increase in the gravitational strength with increasing energy. Ascribing it to a a running $G$ would yield Eq. 2, or equivalently the momentum factor in square brackets in Eq. \ref{potential}. Note that the parameter $\mu$ does not enter the coordinate space potential at finite $r$ because the Fourier transform of a constant is a delta function. We keep $\mu \sim M_P$ for the low energy validity of the effective theory.

The scattering amplitude leading to the result in Eq. \ref{potential} includes all diagrams, including box and crossed-box diagrams and some triangle diagrams. In QED or QCD we do not use the full set of diagrams for the running charge, as we include only the vertex corrections and vacuum polarization. In these theories, this is appropriate because it is these diagrams that renormalize the gauge charge. In gravity, none of these diagrams renormalize G at low energy, so the rationale for including only a subset is not clear. Moreover, in gravity, this subset of diagrams does not by itself form a gauge invariant set. Nevertheless, we can look at this subset of diagrams in a particular gauge. In harmonic gauge, the inclusion of both vertex and vacuum polarization would be
\begin{equation}
G(q) =G \left[1- \frac{167}{60\pi}G{\mathbf{q}}^2\ln \left( \frac{\mu^2}{{\mathbf q}^2}\right)\right]\,,
\end{equation}
i.e. it carries the opposite sign from the full result.
Even within this subset, one potentially might like to exclude the vertex diagrams because there is no Ward identity that indicates that these must be the same for all
particles (i.e. photons vs planets). The vacuum polarization is however universal. Including only it would yield
\begin{equation}
G(q) =G \left[1+ \frac{43}{60\pi}G{\mathbf{q}}^2\ln \left( \frac{\mu^2}{{\mathbf q}^2}\right)\right]\,,
\label{purevacpol}
\end{equation}
as can be seen from Sec. 4.
Overall, the non-relativistic scattering amplitude is made up of many large contributions that differ in sign and magnitude. Identifying a subset as the running charge would not capture the leading quantum effects. Moreover, we see that even the sign of the potential definition is not obvious as vertex and vacuum polarization diagrams have opposite signs. The sign of vacuum polarization correction in Eq. \ref{purevacpol} agrees with that of the the total scattering amplitude in Eq. \ref{potential}, but the magnitude is different by a factor of three.

\section{Lessons}

We have explored one-loop calculations in general relativity in the region where there is perturbative control over the theory. There emerges no definition of a running $G$ that is both useful and universal. The nature of the energy expansion of the effective field theory of gravity implies that quantum corrections are associated with the renormalization of higher order operators rather than the original Einstein action. This implies that the usual theoretical framework for running couplings, the renormalization group, does not by itself define a running $G$. And while a definition generally can be made that is useful within a given process, the quantum effects are so non-universal that this definition will not be usefully applied to other reactions.

We have illustrated a series of reactions with one-loop corrections which differ significantly. Perhaps lost in the variety of signs and magnitudes is the key point that quantum corrections do not organize themselves into a running coupling. This is the expected behavior of an effective field theory. The relevant higher order operators are process dependent and decoupled from the renormalization of the lowest order operator. The kinematic variation of the one loop corrections are more complex than just mirroring the leading behavior because they involve higher powers of the momentum invariants and there are many allowed kinematic factors present at higher order. Attempts to repackage this larger kinematic variation as if it were a modification of the lowest order amplitude, i.e. a running coupling, will then in general fail because the running coupling cannot mimic the richer kinematics of the higher order terms. This leads directly to the crossing problem and the non-universality problem, both of which occur when one tries to define a running $G$.

Our work also provides cautions for the Asymptotic Safety program, which employs a running gravitational coupling in the non-perturbative regime beyond the Planck scale. Let us mention some of the obstacles. The process of defining the running coupling requires a truncation of the operator basis, and the effects of the infinite set of higher order operators get repackaged as if they were contained in a small set of low order operators. This raises the issues that appears in our calculations - will this repackaging be universal? Will the matter couplings in the theory - which provide one definition of $G$ - have the same quantum corrections as the pure gravity sector - which provide another definition of $G$? Will 2-point, 4-point  and 8-point functions, for example, have the same behavior? Due to the presence of all the higher order operators, general covariance by itself does not require these functions to have the universal behavior expected from a running coupling constant. Our calculations showed highly non-universal behavior. Another issue is the continuation back to physical Lorentzian spacetime. In Asymptotic Safety, the running coupling defined in Euclidean space must be continued to physical spacteimes when applied to the real world. Lorentzian spacetimes have momentum variables of both signs and the analytic continuation of power corrections must be more complicated, and we have seen examples of reactions where the crossing problem made any running coupling useless. A naive continuation of a function such as Eq. \ref{runningform} raises the possibility of poles for certain kinematic configurations.

The best candidate for a universal contribution to a running $G$ comes from the vacuum polarization amplitude. The correction to the graviton propagator will occur anytime a graviton is exchanged, either at tree level or within loops. However, here the perturbative result has the wrong sign when Euclideanized - the gravitational strength increases. There is also the crossing problem in Lorentzian spacetime. Even if the vacuum polarization could be summed to a function with a good high energy behavior, for example a form such as
\begin{equation}
\frac{{\cal P}^{\alpha\beta\gamma\delta}}{q^2 + \alpha G q^4}~~,
\end{equation}
but such functions often have trouble with ghosts when used at high energy or within loop diagrams.
\footnote{ For other challenges with the trans-Planckian gravity see e.g. \cite{Dvali:2010ue} and \cite{Giddings:2010pp}. }
The potential problems of gravity treated beyond the Planck scale need not be problematic if the effective theory gets modified at that scale by new degrees of freedom and a change in the description of the theory. General relativity would still form a quantum effective field theory with calculable quantum effects below the Planck energy. However, in such effective field theories with a dimensional coupling it has not proven useful to employ running coupling constants. We have shown the difficulties of trying to define a running $G$ in gravity.

\section*{Acknowledgements}

The work of J.D. is supported in part by the U.S. NSF grant PHY-0855119. The work of  M.A. has been supported by NSERC Discovery Grant of Canada. John Donoghue acknowledges the kind hospitality of the Niels Bohr International Academy, where most of this research was accomplished, and thanks N.E.J. Bjerrum-Bohr for useful conversations.

\end{document}